# CLUSTER PRODUCTION IN QUARK-HADRON PHASE TRANSITION IN HEAVY-ION COLLISIONS

**Rudolph C. Hwa**[1], **C. S. Lam**[2], and **Jicai Pan**[1,2]

[1]Institute of Theoretical Science and Department of Physics
University of Oregon, Eugene, Oregon 97403

[2] Department of Physics, McGill University
Montreal, Quebec, Canada H3A 2T8

## Abstract

The problem of cluster formation and growth in first-order quark-hadron phase transition in heavy-ion collisions is considered. Behaving as Brownian particles, the clusters carry out random walks and can encounter one another, leading to coalescence and breakup. A simulation of the process in cellular automaton suggests the possibility of a scaling distribution in the cluster sizes. The experimental determination of the cluster-size distribution is urged as a means to find a clear signature of phase transition.

The JACEE cosmic-ray data [1,2] on rapidity distribution of hadron production show large fluctuations of multiplicity from bin to bin. Hydrodynamical study of high-energy nuclear collisions can at best describe only the average thermodynamical quantities, relinquishing any possibility of addressing the issue of fluctuations. Event generators incorporating hard scattering of partons in addition to conventional soft production of hadrons, after summing over the many nucleon-nucleon collisions involved, show very little fluctuation of non-statistical nature [3-5]. In our view phase transition is the only likely source of large fluctuations. We consider here a possible dynamical origin for fluctuations in heavy-ion collisions. Our focus is on first-order quark-hadron phase transition (PT). We describe a mechanism in which large clusters of hadrons can be produced as a result of the PT. Furthermore, we suggest what should be investigated experimentally to identify interesting features of cluster production.

It is necessary to state from the outset that any hadronic signature from the PT would be washed out, if there is a hadron gas in thermal equilibrium shielding the plasma undergoing PT and randomizing any signal that might otherwise reach the detector. Thus we assume the plasma break-up scenario in which the hadrons are ejected from the boundary of the mixed phase and fly off in free expansion without further interaction and thermalization, a view that has recently acquired phenomenological interest even at present energies [6,7]. If the data on hadron production in heavy-ion collisions at even higher energies show evidences of clustering, they will by themselves suggest the absence of a hadron gas, or at least its ineffectiveness in damping the hadronic signature.

If the PT is of second order, conventional theory in condensed-matter physics would suggest the existence of long-range correlations and the associated fluctuations. The application of such ideas to nuclear collisions has been carried out already in previous studies that resulted in the identification of certain observables related to intermittency [8]. For first-order PT the nature of the problem is different, so the observables will be different. There is at this point no suggestion from first principles what the optimal observables should be to reveal the effects of such a PT in heavy-ion collisions. Lattice QCD is a microscopic theory implemented in a way too coarse to make possible any prediction on the nature of hadron multiplicity fluctuations. Hydrodynamical studies are too macroscopic and smooth to be concerned about fluctuations. What is needed is something in between. That is our problem here.

We present in this note a nontraditional approach to the PT problem. We shall describe a cellular automaton that simulates the production of hadronic clusters. A set of simple rules will be adopted to represent the dynamical process of cluster growth and the time evolution of the plasma. The aim is to determine the distribution of cluster sizes.

First, let us state why we depart from the traditional approach. An important feature in the first-order PT is that there is a mixed region of quarks and hadrons. Those hadrons are not the same as the ones that end up in the detector, since they are at finite temperature $T_c$. Furthermore, the hadrons in the mixed region are surrounded by quarks (and gluons, which we omit mentioning for brevity), which can make the hadrons grow in size, as they themselves hadronize. The growth process can depend on many factors. In cosmic PT the hadronic bubble can grow to a very large size, the dynamics being determined by the surface tension, the pressure difference at the quark-hadron interface and other factors. Indeed, the bubbles can become so large compared to the range of microscopic interaction that one can use hydrodynamics to treat the space-time behavior of the quark-hadron boundary [9,10]. In heavy-ion collisions the bubbles cannot become very large, since the region of the mixed phase can at most be several fm thick, not more than 10 fm, say. The time scale involved is also short, so a bubble cannot grow beyond a few fm in radius before it is thrust out of the plasma system. Thus all scales in the problem, e.g., QCD scale, hadron mass, surface tension, bubble size, etc., are of the same order of magnitude. In that case one cannot justify hydrodynamical treatment of the bubble boundary and the conventional approach to the determination of the nucleation rate based on the free-energy difference at the quark-hadron interface [11]. Since it becomes necessary to consider, in addition to the usual volume and surface terms, the curvature term (linear in the radius) as well as other terms that become relevant if the bubble is not spherical [12], the complexity of the problem seems to get out of hand.

In the face of such complications, one can either ignore them temporarily and proceed in the conventional way to see what emerges, or take a very different tack. Instead of calculating the nucleation rate and the plasma expansion rate [13], which are important to know, we redirect our questions to ask not so much what the average value of some dynamical quantity is, but what the fluctuation of some observable from the average may be. That may seem like being over-ambitious, but we have a rather modest goal that is not unrealistic. We ask whether there are aspects about the dynamics of bubble growth that affect the fluctuations of the bubble sizes in ways that are tractable. If the average value of the sizes depend on so much detail that simplified calculation is not likely to yield reliable result, then normalized moments of the fluctuations may be relatively free of that defect. Of more importance is to discover possible scaling behavior of the fluctuations, which may convey the essence of the dynamics independent of the details. We describe below an approach to the problem that stands between the microscopic and hydrodynamical approaches.

We first remark that since the hadronic volume in the mixed phase need not be spherical, it can have irregular shapes like dendrites with fingers [12]. We shall therefore refer to them hereafter as clusters, instead of bubbles. They need not have rigid structures as they grow, so they are not of the type usually studied in local growth models or diffusion-limited growth models [14]. Our focus is on the dynamical mechanisms that affect the growth process. The most important mechanism is obviously the inelastic collision between clusters that leads to coalescence. This includes nucleation near the boundary of an existing cluster, which then becomes larger in the traditional sense of growth. In our description of growth through coalescence it can occur between two encountering clusters even if the system is not in equilibrium, for which the use of free energy would be meaningless. Moreover, the breakup of a large cluster into smaller ones due to collisions is also possible, such as in the severing of a finger from a dentrite. Thus if one were to write a stochastic equation describing the evolution of the cluster-size distribution function, it should have both gain and loss terms. Between steps of coalescence and breakup, a cluster moves as a color neutral subsystem that is under the influence of collisions by quarks and gluons in the mixed phase. Each cluster should therefore undergo Brownian motion, which leads to random relative velocities among the clusters when they collide. That in turn can possibly influence the coalescence and breakup rates. The problem seems untractable analytically.

The problem is analogous to that of many random walkers, generated randomly at all points in space and time in the mixed region with some fixed initial size; they can increase or decrease in size, or die, as walkers meet. We put these ideas in a cellular automaton to simulate the cluster growth problem. The spirit of the calculation is to adopt simple rules that capture the essence of the dynamics of growth and see whether there exists in the result some universal features that transcend the severe approximations made on the details.

The physical problem is that of the collision of two high-$A$ nuclei at very high energy, say, at RHIC or even better at LHC. After impact there is rapid longitudinal expansion with slower radial expansion. To simplify the problem we restrict our attention to a thin slice of the expanding cylinder at midrapidity. If temperature $T$ is a good notion, then there is a $T$ profile, with $T$ higher in the interior, lower near the boundary of the disc under consideration. Let there be an annular ring in which the plasma is in the mixed phase at $T_c$. In some hydrodynamical solution of the problem that ring can in a short time become the whole disc and then shrink in radius [15]. We do not rely on the validity of hydrodynamics to proceed. Making use of the azimuthal symmetry of the geometry, let us map the mixed region to a 1-dimensional interval between $r = 0$ and $L$ at an initial time $t = 0$ when hadrons begin to form, and consider the problem of cluster formation in this 1-d space, for simplicity. The M region, as we shall call it, has a fixed boundary at 0 and a moving outer boundary at $r_b(t)$, where $r_b(0) = L$. The dependence of $r_b$ on $t$ will be a result of the process of cluster formation and emission from the plasma. When $r_b(t) = 0$, the phase transition is completed.

To construct a cellular automaton, we discretize space and time and set up some rules for computation. The rules will embody the dynamics of the cluster formation process, which we discuss as we proceed.

*Space-time.* The spatial coordinate is discretized into sites separated by units of $l$ with $L = 100l$. Each incremental time step is taken to be equal to some unit $\tau$, whose exact value is immaterial, since it will be scaled out.

*Nucleation.* We summarize the detailed dynamics of nucleation by a probability $p$ that a cluster (hadron, initially) can be formed at any given site in the M region. The initial cluster size $S_0$ is set to be $l$, so that two neighboring sites can both possibly be the nucleation centers with the "clusters" on them being regarded as barely touching. If we think of a nucleated site as occupied, then at the next time step only the unoccupied sites can become occupied with probability $p$ at each site again. If one worries about supercooling and latent heat release, the value of $p$ may change. In particular, one may want to consider the possibility that $p$ is somewhat larger at a site adjacent to an existing cluster, whose size is larger than a certain value, to simulate the enhancement of nucleation on the surface, if that is the physics one regards as pertinent and wishes to incorporate. We simplify the problem here by adopting an average constant value of $p$ at all unoccupied sites at all times for the whole process.

*Average Drift.* Because of the radial expansion, there is an average drift of all quarks, gluons, and hadrons toward the outer boundary. We represent that motion in our M region by requiring all occupied sites to move in the next time step to, on average, one site outward. That is, if at $t_1$ a cluster is centered at site $s_1$, whatever its size, then at $t_2 = t_1 + 1$ the cluster moves to $s_1 + 1$ (on average), all space (time) increments being in units of $l$ ($\tau$). The scale of $\tau$ is chosen to render the proper drift velocity.

*Boundaries.* In the more realistic scenario the quark phase on the inside of the annular ring mentioned earlier supplies the quarks to the mixed region, while the hadron clusters leave the ring on the other side. The quarks never leave the outer boundary due to confinement. They leave only after converting to hadrons. In our 1-d model that is represented by a fixed boundary at $r = 0$ with undiminished quark density despite the drift of all matter outward, i.e., $r = 0$ is always an unoccupied site available for nucleation at every new time step. That is the source. The sink is the ejection of hadronic clusters at $r_b$, which is defined to be the outermost unoccupied site until it is occupied. When a cluster of size $S$ reaches the boundary $r_b(t)$ at $t$, it is removed from the M region and at the next time step $r_b(t+1) = r_b(t) - S$. Thus the mixed region is reduced in extension as clusters leave it. When all quarks in M are hadronized, $r_b$ becomes 0 and the phase transition is over.

*Cluster-Size Distribution.* Each cluster that leaves the boundary contributes to the cluster-size distribution. Whether in the $i$th event in an experiment or in the $i$th simulation of an event, let $n_i(S)$ be the number of clusters of size $S$ after all emerging clusters are collected for the event, and $N_i$ be the total number of clusters of all sizes in the $i$th event, i.e., $N_i = \int_0^\infty dS\, n_i(S)$. Then $P(S)$ is the average distribution after $\mathcal{N}_{\text{evt}}$ events, defined by

$$P(S) = \frac{1}{\mathcal{N}_{\text{evt}}} \sum_{i=1}^{\mathcal{N}_{\text{evt}}} \frac{n_i(S)}{N_i}. \tag{1}$$

The determination of $P(S)$ is the aim of this calculation.

*Random Walk.* Since the clusters in the plasma behave as massive colloids in a fluid, they undergo Brownian motion, which we simulate by requiring them to take random walks around their average drift. Thus each cluster is assigned at the next time step two possible sites with equal probability around the average drift site, i.e., if at $t_1$ a cluster is at site $s_1$, then at $t_2 = t_1 + 1$ the cluster moves to either $s_1$ or $s_1 + 2$.

*Collisions.* All clusters have integer sizes in units of $l$. Let a cluster $C_i$ centered at site $s_i$ have size $S_i$, which may be large enough so that $C_i$ may occupy a few neigboring sites of $s_i$. None of those covered sites are allowed to nucleate further. As clusters take their steps in random walk, they may encounter each other. A collision is defined to take place between two encountering clusters when they overlap at least half a unit in $l$. For example, two clusters with $S_i = S_j = 2$ and $|s_i - s_j| = 2$ are not in collision, but if $S_i = 3$, the other variables unchanged, then they are in collision. When a collision occurs at $t$, we consider three possible outcomes. In each case we rearrange the products at $t$ as initial positions for evolution to the next time step.

*Coalescence:* $C_i + C_j \to C_k$. We require that the size increases additively: $S_k = S_i + S_j$. $C_k$ is placed at the center of mass of $S_i$ and $S_j$. If that is not a regular site position on the lattice, then let $s_k(t)$ be the site closest to the position $[S_i s_i(t) + S_j s_j(t)]/(S_i + S_j)$. We use $c$ to denote the probability for coalescence when two clusters collide.

*Breakup:* $C_i + C_j \to C_i + C_k + C_l$. We assume that either $C_i$ or $C_j$ may break up, not both, and never one with the smallest possible size $l$. The site for $C_i$ (the unbroken one indicated above) is unchanged from $s_i(t)$; the broken pieces $C_k$ and $C_l$ are given random partitioning probability subject to the constraint $S_k + S_l = S_j$. They are assigned the sites $s_k(t)$ and $s_l(t)$ closest to the positions $s_j(t) \pm S_l(t)/2$ and $s_j(t) \mp S_k(t)/2$, respectively, where $\pm$ are randomly determined. We use $b$ to denote the probability for breakup when two clusters collide.

*Elastic:* $C_i + C_j \to C_i + C_j$. When a collision does not lead to either coalescence or breakup, we let the two clusters scatter elastically. Their positions are unchanged. Without randomness the next step can be either forward or backward scattering. But the rule in random walk determines the positions at the next time step.

*Evolution.* After the collisional outcome is determined at $t$, we further allow the medium to generate new clusters of size $l$ with probability $p$ at each unoccupied site such that no new clusters can overlap with the existing ones. Then we let each of the clusters now at $t$ to take a step in random walk, and repeat the above process. With these rules the system will evolve in time until all clusters are ejected from the plasma at $r_b(t)$ with varying sizes, yielding $n_i(S)$ for the $i$th event.

We note that the rules given above do not depend on the system being in thermal equilibrium, and no equation of state is assumed. They embody the skeleton of the dynamical process that we regard as most important in the conversion of quarks into hadrons in the mixed region. They stand between the microscopic and macroscopic descriptions mentioned earlier. The rules can be modified to incorporate other factors deemed important later. The purpose at this point is not to find the perfect set of rules, but to see whether some reasonable rules will lead to something interesting that may be relevant phenomenologically.

With the above rules, we have simulated the PT process 20K times. We set $p = 0.05$ and obtained $P(S)$ for various values of the coalescence probability $c$ and breakup probability $b$. The results are shown in Fig. 1(a) and (b) in log-log plots. Evidently, there is linear behavior in all cases shown until finite-size effect cuts off the approximate linearity. For the linear portion one may express $P(S)$ as possessing a power-law behavior

$$P(S) \propto S^{-\sigma}. \qquad (2)$$

The scaling exponent $\sigma$ depends on the dimensionless parameters $b$ and $c$ in the expected way, i.e., $\sigma$ increases with $b$ but decreases with $c$. In Fig. 1(c) we show the dependence of $P(S)$ on $p$ for some typical values of $b$ and $c$. It is evident that there is approximate independence on $p$ until $p$ exceeds 0.05. At higher $p$ the rate of cluster production is higher, thereby shortening the lifetime of the M region, which in turn suppresses the development of large clusters. Nevertheless, before the finite-size effect becomes important the presence of a $p$-independent scaling region is interesting.

The point of this calculation is to show that the dynamics of cluster formation and growth can lead to the possibility of a scaling result in $S$. The discovery of any value of $\sigma$ that is not very large would be an indication of cluster growth during PT and therefore exciting. It is hard to think of any competing dynamical process that would lead to the production of large clusters. Thus, a nontrivial behavior such as (2) can be a distinctive and unambiguous signature of PT; it indirectly implies that the system has first been in the state of a quark-gluon plasma. Of course, it is the experimental discovery of any such scaling behavior that is of real significance. Here we can merely suggest what should be looked for in an experiment, and our simulation based on the essential elements of the dynamics provides some concrete result to motivate such a search.

Experimentally, it is necessary to analyze the event structure in the three dimensional space $(\eta, \phi, \ln p_T)$. Since cluster formation at different times may have different transverse momenta (which, for example, is undoubtedly $\approx 0$ at the end of PT), it is important to avoid the superimposition of the various clusters produced by not integrating over all $p_T$. One can make various cuts in $p_T$ so that in each small range of $p_T$ one examines the multiplicity distribution in $(\eta, \phi, p_T)$. A cluster is to be identified as an island in such a space, surrounded by empty cells (or bins). The cluster size $S$ is then the number of particles in an island. Note that such a definition of cluster depends on the resolution scale, since at high enough resolution every particle is by definition an island of its own. On the other hand, if the resolution is too coarse, all particles form one big cluster. Clearly, an appropriate variation of the bin size is necessary in order to identify cluster structure.

If one wishes to emphasize the fluctuations rather than the average size $\langle S \rangle$ as being the more interesting measure of the dynamical process, one can calculate the normalized moments

$$C_q = \frac{\langle S^q \rangle}{\langle S \rangle^q}, \qquad (3)$$

where $\langle \cdots \rangle$ denotes averaging by $P(S)$. Unlike intermittency [2], one should not expect power-law behavior of $C_q$ as a function of the resolution size. Nevertheless, characteristic behavior corresponding to (2) would show up, if at all, only for certain bin sizes. One should therefore investigate $C_q$ as a function of the cell size in the 3-dimensional phase space for various sectors of $E_T$ and $p_T$. Any nontrivial $q$ dependence of $C_q$ would be very interesting.

It is of interest to point out that the scaling behavior (2) would hint at the possibility of self-organized criticality [16] for our cluster production process. Indeed, there is a similarity between cluster growth and the sandpile problem [16,17], where the produced cluster is the analogue of an avalanche from the sandpile. In that problem when the sandpile reaches a critical slope, a grain of sand can tumble down the slope and either increase in size or dissipate into the pile without inducing an avalanche. In the end what falls off the table can have a distribution of avalanche sizes, satisfying a scaling law similar to (2). In our problem clusters can gain in size in a coalescence, or diminish in size in a breakup. When they emerge from the $M$ region, they have a scaling distribution of sizes. A large size $S$ is rare, but possible. In neither problem is there a fundamental theory reliable enough for analytical calculation; both use cellular automata to simulate what can likely happen based on some simple rules. The sandpile problem illustrates the phenomena that many systems move themselves into a "critical point" without external tuning of parameters. In our problem also equilibrium thermodynamics does not play a pre-eminent role. In both problems a small fluctuation in the evolution process can lead to large-scale fluctuation in the outcome at little cost in energy. It is possible that our conventional notion of what happens at phase transition for systems in equilibrium is inappropriate for high-energy nuclear collisions; the uncertainty is made worse by the fact that we have no experimental way to directly prove the validity of that notion. It is further possible that the system may be driven by a dynamics that leads itself to self-organized criticality without thermal equilibrium. If that can be found to be closer to reality, then we would be at the threshold of discovering something very unconventional and interesting. The possibility that cluster production in heavy-ion collisions can convey some hint of such a phenomenon is therefore intriguing.

To summarize, we have considered the problem of cluster growth in heavy-ion collisions when there is a quark-hadron PT, and demonstrated by use of a cellular automaton that a scaling distribution of cluster sizes is a likely phenomenological possibility. We strongly urge the experimental measurement of clusters to check our suggestion. The observation of any clustering of produced hadrons with a nontrivial distribution of their sizes would be an unambiguous signature of something unusual and stimulating, and would reveal the properties of cluster formation that cannot be probed any other way.

We have benefitted from helpful discussions with Per Bak, Martin Grant, Hong Guo, Terry Hwa and Ben Svetitsky, to whom we are grateful. This work was supported, in part, by the U.S. Department of Energy under Grant No. DE-FG06-91ER40637, and by the Natural Sciences and Engineering Council of Canada and the Quebec Department of Education.

**Figure Caption**

Fig. 1  (a) and (b) Distributions of cluster sizes for various values of coalescence probability $c$ and breakup probability $b$ as indicated; (c) $P(S)$ for various values of initial formation probability $p$.

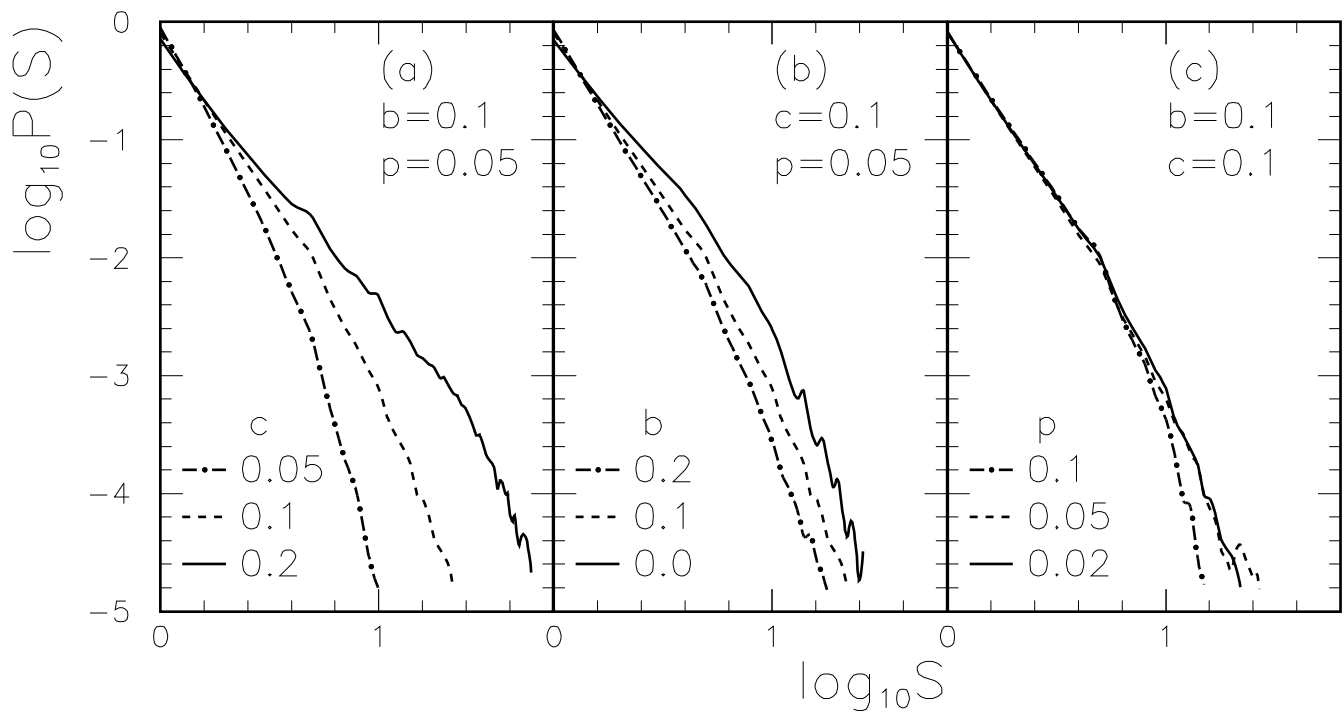